\documentclass[twocolumn,showpacs,preprintnumbers,prb,amsmath,amssymb]{revtex4}
\makeatletter
\def\@dotsep{4,5}
\makeatother

\usepackage{graphicx,color} 
\usepackage{dcolumn}
\usepackage{bm}
\usepackage{subfigure} 

\begin{document}


\title{Minimum free-energy path of homogenous nucleation from the phase-field equation}

\author{Masao Iwamatsu}
\email{iwamatsu@ph.ns.tcu.ac.jp}
\affiliation{
Department of Physics, School of Liberal Arts,
Tokyo City University (formerly Musashi Institute of Technology),
Setagaya-ku, Tokyo 158-8557, Japan
}%


\date{\today}

\begin{abstract}
The minimum free-energy path (MFEP) is the most probable route of the nucleation process on the multidimensional free-energy surface.  In this study, the phase-field equation is used as a mathematical tool to deduce the minimum free-energy path (MFEP) of homogeneous nucleation.  We use a simple square-gradient free-energy functional with a quartic local free-energy function as an example and study the time evolution of a single nucleus placed within a metastable environment.  The time integration of the phase-field equation is performed using the numerically efficient cell-dynamics method.  By monitoring the evolution of the size of the nucleus and the free energy of the system simultaneously, we can easily deduce the free-energy barrier as a function of the size of the sub- and the super-critical nucleus along the MFEP. 
\end{abstract}

\maketitle

\section{\label{sec:sec1}Introduction}
\label{sec:level1}
Nucleation is a very basic and ubiquitous phenomena that occurs in phase transformation of various materials~\cite{Oxtoby1992}.  A small embryo of new stable material appears from the fluctuation within a metastable material, and it grows into a small cluster called a nucleus.  Then it overcomes the free-energy barrier and will grow indefinitely.  The basic process of the formation of a single nucleus called nucleation is difficult to study even theoretically as the process involves non-equilibrium and transient states.  In much of the literature~\cite{Oxtoby1988,Zeng1991,Iwamatsu1993} only the critical nucleus that corresponds to the cluster at the top of the energy barrier has been considered and used to deduce the activation energy of the nucleation rate.  

The whole landscape of the free-energy surface for a single nucleus is impossible to study because of the huge degree of freedom even for a small cluster.  Intuitively, however, it is believed that the sub-critical nucleus ascends the free-energy surface along the valley to the top of the barrier.  The top of the barrier is in fact the saddle point of the free-energy surface.  The nucleus at the saddle point is called the critical nucleus.  It surpasses the barrier at the saddle point and descends along the valley as a super-critical nucleus and will grow indefinitely. This minimum free-energy path (MFEP) is the most probable route of the reaction for the nucleation even though the real path might wander around this MFEP through a thermal fluctuation or some other effect.

In order to extract information about the free-energy surface for the sub- and the super-critical nucleus, Weakliem and Reiss~\cite{Weakliem1993} proposed a theoretical model called the molecular theory of nucleation.  In this theory, the $(N,\lambda)$ cluster model where a fixed number of molecules $N$ confined within a container of the radius $\lambda$ (or the volume $V$) are considered.  Using a physically plausible boundary condition one can study the projection of the free-energy surface onto the two-dimensional $(N,\lambda)$  (or $(N,V)$) space which is expected to capture the essential features of the real free-energy surface in multi-dimensional space.  This theory was originally formulated using the Monte Carlo method and was reformulated using the density-functional theory (DFT)~\cite{Talanquer1994,Uline2007}.  However, it is not obvious how to restore this projected free-energy surface in the two-dimensional $(N,\lambda)$ space to the true free-energy surface and barrier in the multi-dimensional space.  There are also many problems inherent to the confined system~\cite{Reguera2003}.  Another theoretical model to use the pressure $P$ instead of the virtual radius $\lambda$ has also been studied~\cite{Gunther2003}.  

The simplest way to study this free-energy surface is to look at only the MFEP for a single nucleus as the function of the number of molecules $N$ as a reaction coordinate.  In this case, the fluctuation of the size $\lambda$ or the volume $V$ for the given number of molecules $N$ considered in the $(N,\lambda)$ cluster model of the molecular theory~\cite{Weakliem1993, Talanquer1994,Uline2007} is neglected.  The Monte Carlo simulation with biased sampling~\cite{Wolde1998,Auer2001} can be used to trace the MFEP. Similarly, by choosing the parameterized spherical density profile, or by using constraints in the minimization process we can calculate the MFEP as the function of the number of molecules $N$~\cite{Lutsko2008a,Lutsko2008b} using DFT.  

In this paper, we will propose an alternative novel method to deduce the MFEP using a much simpler phase-field equation~\cite{Pusztai2008} for the non-conserved order parameter.  Since the time integration of the highly non-linear phase-field equation requires a certain amount of computational resources~\cite{Pusztai2008}, we will further simplify the equation using the cell-dynamics method~\cite{Iwamatsu2008a}.  The format of this paper is as follows: Section \ref{sec:sec2} is a brief review of the phase-field equation and the free-energy model used to study the MFEP. In section \ref{sec:sec3}, the cell-dynamics method is used to integrate the phase-field equation and the MFEP is extracted from the time evolution of the size and the free energy of a single nucleus.  We will conclude this paper with our final comments in section \ref{sec:sec4}.

\section{Phase-Field Equation and the Minimum Free-Energy Path}
\label{sec:sec2}

In order to study the evolution or the regression of a single nucleus~\cite{Bagdassarian1994,Wild1997} and the reaction-path of nucleation called the minimum free-energy path (MFEP), we will use the partial differential equation called the phase-field equation for the non-conserved order parameter:
\begin{equation}
\tau\frac{\partial \psi}{\partial t}=-\frac{\delta F}{\delta \psi}
\label{eq:1-1}
\end{equation}
where $\psi$ is the order parameter called phase field, $F$ is the free-energy functional (grand potential) usually expressed by the square-gradient form~\cite{Iwamatsu1993,Bagdassarian1994,Wild1997}, and $\tau$ specifies the time scale of evolution.  This phase-field equation is attractive as the kinetics of phase transformation is driven by the relative stability of each phase and by the topology of the associated free-energy surface $F$.  In particular, the special solution of Eq.~(\ref{eq:1-1}) is the interface-controlled growth of a spherical or a circular grain with nearly constant velocity~\cite{Chan1977} in $d=3$ and $d=2$ dimensions respectively.  The dynamics driven by Eq.~(\ref{eq:1-1}) is also attractive to study MFEP as it always guarantees that the total free energy decreases monotonically~\cite{Langer1992}:
\begin{equation}
\frac{dF}{dt}
=\int \frac{\delta F}{\delta \psi}\frac{\partial \psi}{\partial t}dr
=-\int\left(\frac{\delta F}{\delta \psi}\right)^{2}dr\leq 0.
\label{eq:1-3}
\end{equation}
Mathematically, the evolution of the nucleation is the trajectory of the dynamical system in infinite dimensions described by Eq.~(\ref{eq:1-1}) in the phase space $\psi$ whose Lyapunov functional is the free energy $F$.   

Since Eq.~(\ref{eq:1-3}) implies the steepest descent of the trajectory of $\psi$ which represents a single spherical or a circular nucleus on the free-energy surface $F$, the phase-field kinetics equation Eq.~(\ref{eq:1-1}) for the non-conserved system can be used as a {\it mathematical tool} to study the MFEP of nucleation not only for the {\it non-conserved} order parameter~\cite{Bagdassarian1994} but also for the {\it conserved} order parameter~\cite{Wild1997}.  Equation (\ref{eq:1-1}) can also be interpreted as an over-damped dynamics on the potential surface $F$ in the functional space $\psi$.  In fact, our use of Eq.~(\ref{eq:1-1}) to explore the MFEP is analogous to the over damped Langevin dynamics used in the string method proposed by Qiu et al.~\cite{Qiu2008} to study the transition pathway of nucleation in the capillary condensation.

For the conserved order parameter, however, the kinetic equation
\begin{equation}
\tau\frac{\partial \psi}{\partial t}=\nabla^{2}\left(\frac{\delta F}{\delta \psi}
\right)
\label{eq:1-3a}
\end{equation}
should be used instead of Eq.~(\ref{eq:1-1}).  Also, in the more rigorously formulated dynamical density-functional theory (dynamical DFT)~\cite{Marconi1999,Archer2004}, the kinetic equation is given by
\begin{equation}
\frac{\partial \rho}{\partial t}=\nabla\left(\rho \nabla\frac{\delta F}{\delta \rho}
\right),
\label{eq:1-3b}
\end{equation}
where the order parameter $\rho$ represents the density. These kinetic equations Eqs.~(\ref{eq:1-3a}) and (\ref{eq:1-3b}) could also be used to search for the free-energy surface as they can also satisfy the inequality similar to Eq.~(\ref{eq:1-3}) which guarantees a monotonically decreasing free energy as a function of time.   

To be exact, the real physics of the nucleation in a conserved system will be described by Eqs.~(\ref{eq:1-3a}) or (\ref{eq:1-3b}).  However, these equations have a shortcoming in that the dynamics will be complex due to the depletion effect as the integrated order parameter or the total density must be conserved.  Also, the mathematical structure of Eqs.~(\ref{eq:1-3a}) or (\ref{eq:1-3b}) implies that they do not trace the free-energy surface $F$ itself in a straight forward maner.  In contrast, the phase-field equation (\ref{eq:1-1}) can trace the free-energy surface and the MFEP directly.  Therefore, we will use Eq.~(\ref{eq:1-1}) to study the MFEP even if it may not have a physical meaning for the nucleation in conserved systems such as liquid-vapor nucleation.  

\begin{figure}[htbp]
\begin{center}
\includegraphics[width=0.8\linewidth]{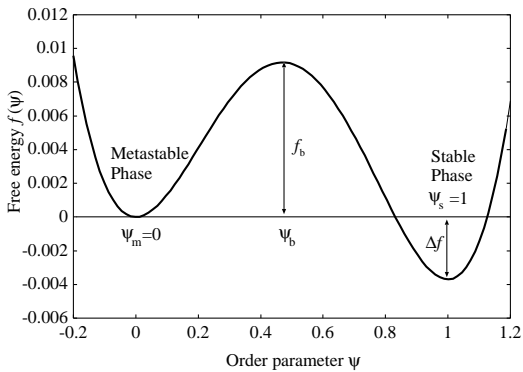}
\caption{
The local part of the free energy Eq.~(\ref{eq:1-4}) when $\eta=0.7$ and $\epsilon=1/75\simeq 0.01333$.  We consider $\epsilon>0$ so that the phase at $\psi=1$ is always stable.
}
\label{fig:1}
\end{center}
\end{figure}

As an example, we will employ the square-gradient model for the free-energy functional:
\begin{equation}
F[\psi]=\int \left[\frac{D}{2}(\nabla \psi)^{2}+f(\psi)\right]d{\bf r} 
\label{eq:1-2}
\end{equation}
where the local part of the free energy $f(\psi)$ proposed by Jou and Lusk~\cite{Jou1997} is used:
\begin{equation}
f(\psi) = \frac{1}{4}\eta\psi^{2}(\psi-1)^{2} + \frac{3\epsilon}{2}\left(\frac{\psi^{3}}{3}-\frac{\psi^{2}}{2}\right),
\label{eq:1-4}
\end{equation}
which represents the two-phase system with two phases characterized by $\psi=0$ and $\psi=1$ shown in Fig.~\ref{fig:1}.  The parameter $\epsilon$ controls the relative stability of one phase at $\psi=1$ with the grand potential $f(\psi=1)=-\epsilon/4$ relative to another phase at $\psi=0$ with the grand potential $f(\psi=0)=0$.  We will only consider the case when $\epsilon>0$.  Then the phase at $\psi_{m}=0$ is the metastable and the one at $\psi_{s}=1$ is stable.  The free-energy difference $\Delta f$ (Fig.~\ref{fig:1}) is given by 
\begin{equation}
\Delta f = f(0) - f(1) = \frac{\epsilon}{4}.
\label{eq:1-5}
\end{equation}

The position $\psi_{b}$ of the free-energy barrier is given by
\begin{equation}
\psi_{b}=\frac{1}{2}-\frac{3\epsilon}{2\eta}
\label{eq:1-6}
\end{equation}
and its height (Fig.~\ref{fig:1}) is given by
\begin{equation}
f_{b}=f\left(\psi_{b}\right)=\frac{\left(\eta-3\epsilon\right)^{3}\left(\epsilon+\eta\right)}{64\eta^{3}}
\label{eq:1-7}
\end{equation}
with $f_{b}\sim \eta/64$ when $\epsilon \ll \eta$. Then the spinodal point defined by $f_{b}=0$ is at
\begin{equation}
\epsilon=\eta/3,
\label{eq:1-8}
\end{equation}
and the binodal is at $\epsilon=0$.  A typical shape of the free-energy function $f(\psi)$ near the binodal is shown in Fig.~\ref{fig:1}.  

Now, the special solution of the phase-field equation Eq.~(\ref{eq:1-1}) for the traveling wave of the form $\psi\left(X\right)$ with $X=r-R(t)$ having a single spherical or a circular shape of the radius $R(t)$ satisfies the differential equation~\cite{Chan1977,Jou1997}
\begin{equation}
D\frac{d^{2}\psi}{dX^{2}}+\tau v \frac{d\psi}{dX}-f^{'}(\psi)=0
\label{eq:1-9}
\end{equation}
where
\begin{equation}
\tau v = \tau \frac{dR}{dt}+D\frac{\Delta}{R}
\label{eq:1-10}
\end{equation}
and $\Delta=0, 1, 2$ for $d=1, 2, 3$ dimensions.  The nucleus grows or shrinks according to
\begin{equation}
\frac{dR}{dt} = v\left(1-\frac{R_{c}}{R}\right)
\label{eq:1-11}
\end{equation}
from Eq.~(\ref{eq:1-10}).  Now the nucleus will grow if the radius $R$ becomes larger than the dynamical critical radius $R_{c}$ given by
\begin{equation}
R_{c}=\frac{D\Delta}{\tau v}
\label{eq:1-12}
\end{equation}
otherwise it will shrink~\cite{Langer1992}. The time evolution of the radius $R(t)$ is given by~\cite{Jou1997}
\begin{equation}
vt=\left(R(t)-R_{0}\right)+R_{c}\ln\left(\frac{R(t)-R_{c}}{R_{0}-R_{c}}\right)
\label{eq:1-13}
\end{equation}
from Eq.~(\ref{eq:1-11}), where $R_{0}$ is the initial radius at $t=0$.  For a sufficiently large nucleus ($R(t)\rightarrow \infty$), the radius $R$ grows with a constant velocity $dR/dt=v$ given by~\cite{Chan1977,Jou1997,Iwamatsu2005a}
\begin{equation}
v=\frac{1}{\tau}\sqrt{\frac{D}{2\eta}}3\epsilon
\label{eq:1-14}
\end{equation}
and the order parameter profile $\psi(X)$ has the form
\begin{equation}
\psi(X)=\frac{1}{1+\exp\left(\frac{1}{2}\sqrt{\frac{2\eta}{D}}X\right)}.
\label{eq:1-15}
\end{equation}

Equation~(\ref{eq:1-9}) can also be integrated directly
\begin{equation}
\int_{-\infty}^{\infty} dX \frac{d}{dX}\left(\frac{1}{2}\frac{d^{2}\psi}{dX^{2}}-f(\psi)\right)=-\tau v \int_{-\infty}^{\infty} \left(\frac{d\psi}{dX}\right)^{2}dX
\label{eq:1-16}
\end{equation}
to give another expression for the velocity $v$ 
\begin{equation}
v=\frac{D\Delta f }{\tau\sigma}
\label{eq:1-17}
\end{equation}
expressed by the surface tension $\sigma$ defined by
\begin{equation}
\sigma = 2\sqrt{\frac{D}{2}}\int_{\psi=0}^{\psi=1}\sqrt{f(\psi)}d\psi
=D\int_{-\infty}^{\infty}\left(\frac{d\psi}{dx}\right)^{2}dx.
\label{eq:1-18}
\end{equation}
The direct integration of Eq.~(\ref{eq:1-18}) with the free energy given by Eq.~(\ref{eq:1-4}) is analytically possible only at the two-phase coexistence $\epsilon=0$, and the surface tension $\sigma$ is gives
\begin{equation}
\sigma = \frac{1}{6}\sqrt{\frac{\eta D}{2}}.
\label{eq:1-19}
\end{equation}
Inserting Eqs.~(\ref{eq:1-5}) and (\ref{eq:1-19}) into Eq.~(\ref{eq:1-17}), we can recover Eq.~(\ref{eq:1-14}).  Now we will simulate the evolution of a single nucleus using the phase-field equation Eq.~(\ref{eq:1-1}).  By monitoring the decreasing free energy of a single nucleus as a function of its radius, we can trace the MFEP for the shrinking sub-critical and the growing super-critical nucleus.  

In the classical nucleation theory (CNT), the minimum free-energy path (MFEP) is given by the work of formation $W_{\rm CNT}$ of the nucleus calculated by assuming a spherical or a circular shape and a sharp interface with the size-independent surface tension $\sigma$: 
\begin{eqnarray}
W_{\rm CNT} &=& -\frac{4\pi}{3}R^{3}\Delta f +4\pi R^{2}\sigma,\;\;\;(d=3), \nonumber \\
        &=& -\pi R^{2}\Delta f + 2\pi R \sigma,\;\;\;(d=2),
\label{eq:1-20}
\end{eqnarray}
for (spherical) $d=3$ and (circular) $d=2$ dimensions.  This is the "minimum" free-energy path within the CNT as we have assumed a spherical or a circular shape of the minimum surface area. Other shapes with a larger surface area and the same volume naturally give a higher free energy as expected from the second term of Eq.~(\ref{eq:1-20}). By maximizing this free energy Eq.~(\ref{eq:1-20}) we can deduce the thermodynamic critical radius $R_{c}$ that corresponds to the saddle point of the free-energy surface for a single nucleus:
\begin{equation}
R_{c}=\frac{\sigma \Delta}{\Delta f},
\label{eq:1-21}
\end{equation}
where the meaning of $\Delta$ is the same as that in Eq.~(\ref{eq:1-10}).  Then the free-energy barrier (activation energy) of nucleation rate is the maximum of the MFEP given by Eq.~(\ref{eq:1-20}):
\begin{eqnarray}
W_{\rm CNT}&=&\frac{16\pi}{3}\frac{\sigma^{3}}{\left(\Delta f\right)^{2}},\;\;\;(d=3), \nonumber \\
&=& \frac{\pi \sigma^{2}}{\Delta f},\;\;\;(d=2).
\label{eq:1-22}
\end{eqnarray}
The thermodynamic critical radius of the CNT in Eq.~(\ref{eq:1-21}) is the same as the dynamical critical radius in Eq.~(\ref{eq:1-12}) since both expressions reduce to
\begin{equation}
R_{c}=\frac{\sqrt{2\eta D}}{3\epsilon}
\label{eq:1-23}
\end{equation}
for $\Delta=1$ ($d=2$ dimension), for example, from Eqs.~(\ref{eq:1-5}), (\ref{eq:1-14}) and (\ref{eq:1-19}).

\section{Results of numerical simulation using cell dynamics}
\label{sec:sec3}

\subsection{Cell-dynamics simulation}

Since Eq.~(\ref{eq:1-1}) is a highly non-linear partial differential equation, we will not attempt to solve it directly.  Instead, we follow Oono and Puri~\cite{Oono1988,Puri1988} and transform this partial differential equation into the space-time discretized difference equation called the cell-dynamics equation. Their transformation does not mean an accurate numerical approximation to the original phase-field equation  Eq.~(\ref{eq:1-1}).  Rather, they originally aimed at simulating the global picture of the kinetics of the phase transformation governed by the phase-field equation within the framework of the discrete cellular automata. However, later workers have found that this difference equation can successfully reproduce the time evolution of the original phase-field equation driven by a subtle balance of different phases represented by the free-energy surface $f$~\cite{Ren2001a,Iwamatsu2005a,Iwamatsu2008a}. 

According to the cell-dynamics method, the partial differential equation Eq.~(\ref{eq:1-1}) is transformed into the finite difference equation in space and time of the following form
\begin{equation}
\psi(t+1,n)=M[\psi(t,n)]
\label{eq:1-24}
\end{equation}
where the time $t$ is discrete integer and the space is also discrete and is expressed by the site index (integer) $n$.  We have eliminated the time scale $\tau$ because the cell-dynamics equation  Eq.~(\ref{eq:1-24}) is a coarse-grained approximation to Eq.~(\ref{eq:1-1}) in time and space so that the time scale is irrelevant.  The mapping $M$ is given by
\begin{equation}
M[\psi(t,n)]=g(\psi(t,n))+D\left[<<\psi(t,n)>>-\psi(t,n)\right]
\label{eq:1-25}
\end{equation}
where the definition of $<<*>>$ for the two-dimensional square grid is given by
\begin{equation}
<<\psi(t,n)>>=\frac{1}{6}\sum_{i=\mbox{nn}}\psi(t,i)
+\frac{1}{12}\sum_{i=\mbox{nnn}}\psi(t,i)
\label{eq:1-26}
\end{equation}
with "nn" means the nearest neighbors and "nnn" the next-nearest neighbors of the square grid.  We will only consider the nucleation in $d=2$ dimension. 

Instead of the original hyper-tangential map function $g\left(\psi\right)=\psi-1.3\tanh\psi$~\cite{Oono1988,Puri1988}, we use the map function $g$ that is directly related to the free energy $f(\psi)$ through
\begin{equation}
g\left(\psi\right)=\psi-\frac{df}{d\psi}.
\label{eq:1-27}
\end{equation}
This replacement is essential to study the nucleation and the growth when a subtle balance of the relative stability of two phases realized by the topology of the free-energy surface $f$ plays a crucial role.  

Ren et al.~\cite{Ren2001a} argued that by using the map function like Eq.~(\ref{eq:1-27}) instead of the original hyper-tangential map function one can easily include the effect of asymmetry in the free energy function $f$ and, hence, can take into account the asymmetric character of the two phases within the framework of the cell dynamics.  Subsequently, the present author has demonstrated that by using the map function Eq.~(\ref{eq:1-27}) derived from the free-energy function $f\left(\psi\right)$ one can simulate the growth of a single circular nucleus to confirm the analytical formula~\cite{Chan1977} for the growth velocity~\cite{Iwamatsu2005a}.   Furthermore, by using the map function in Eq.~(\ref{eq:1-27}), the nucleation and the growth process of multiple nuclei can also be successfully simulated not only in the site saturation regime~\cite{Iwamatsu2005a} but also in the continuous nucleation regime~\cite{Iwamatsu2008a} to confirm the KJMA (Kolmogorov-Johnson-Mehl-Avrami) picture of phase transformation\cite{Christian1965}.  It is now well established that this cell-dynamics method can serve as a simple integration scheme to study the kinetics of phase-field model even though the method is not guaranteed~\cite{Teixeira1997} to be an accurate numerical approximation to the original equation Eq.~(\ref{eq:1-1}).  

Since the direct numerical integration of the partial differential equation Eq.~(\ref{eq:1-1}) requires a finely tuned integration scheme~\cite{Bagdassarian1994,Wild1997} and a significant amount of computational resources~\cite{Pusztai2008}, we will use this cell-dynamics equation Eq.~(\ref{eq:1-24}) instead of the original phase-field equation Eq.~(\ref{eq:1-1}) to study the evolution of a single nucleus in this paper.

\subsection{Numerical results}

Initially a circular nucleus with the order parameter $\psi=1$ and the radius $R_{i}$ is placed within the metastable environment with $\psi=0$ in $d=2$ dimensional space.  Then the cell-dynamics equation Eq.~(\ref{eq:1-24}) is solved to simulate the evolution of the initial nucleus. If the initial radius $R_{i}$ is larger than the dynamical critical radius $R_{c}$, the nucleus is expected to grow indefinitely otherwise it is expected to shrink and disappear.  This critical radius $R_{c}$ is also the thermodynamic critical radius of the critical nucleus.  We will use three sets of the free-energy parameters shown in Table~\ref{tab:1}. The parameter $D$ in Eq.~(\ref{eq:1-2}) is set to 1/2 throughout this paper. We use small $\epsilon$ to make the critical radius Eq.~(\ref{eq:1-23}) large.  Therefore we will only consider the thermodynamics state closer to the binodal than the spinodal where $\epsilon=\eta/3$ must be large.

\begin{figure}[htbp]
\begin{center}
\includegraphics[width=0.75\linewidth]{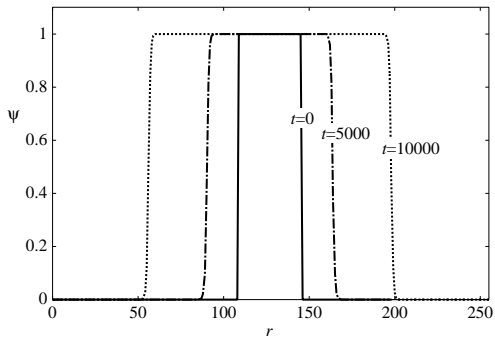}
\caption{
Cross sections of the growing super-critical nucleus when $\eta=1.0$ and $\epsilon=0.01333$. Initially a circular nucleus of $\psi$=1 with a radius $R=18$ and a sharp interface is prepared at the center of 256$\times$256 cells.  
}
\label{fig:2}
\end{center}
\end{figure}

\begin{table}[htbp]
\caption{
Three sets of the free-energy parameters $\eta$ and $\epsilon$ used together with the critical radius $R_{c}$ and the interfacial velocity $v$ from theoretical predictions by Eqs.~(\ref{eq:1-23}) and (\ref{eq:1-14}) and from our cell-dynamics simulations.   }
\label{tab:1}
\begin{center}
\begin{tabular}{c|ccc}
\hline
$\eta$                & 1.0          & 0.7          & 1.0      \\
$\epsilon$            & 0.01333      & 0.01333      & 0.02222 \\ 
\hline
$R_{c}$               & 25           & 21           & 15      \\
$R_{c}$ (Simulation)  & 14$-$17      & 13$-$14      & 8$-$10 \\
$v$                   & $0.020/\tau$ & $0.024/\tau$ & $0.033/\tau$ \\
$v$ (Simulation)      & 0.0070       & 0.0098       & 0.0121 \\
\hline
\end{tabular}
\end{center}
\end{table}

Figure \ref{fig:2} shows the cross sections of the growing super-critical nucleus when $\eta=1.0$ and $\epsilon=0.01333$. Initially a circular nucleus of $\psi$=1 with a radius $R=18$ and a sharp interface is prepared.  Its radius increases and the interface become diffuse in the course of evolution.  The nucleus with initial radius $R=14-17$ neither grows nor shrinks, while the nucleus with initial radius smaller than $R=13$ always shrinks.  The critical radii $R_{c}=14-17$ in Table~\ref{tab:1} estimated from our cell-dynamics simulation are those radii for which the initial nucleus neither grows nor shrinks.  The critical radius are slightly smaller than the theoretical prediction from Eq.~(\ref{eq:1-23}) tabulated in table I that is probably due to the discrete difference Eq.~(\ref{eq:1-26}) used in the cell-dynamics method.

\begin{figure}[htbp]
\begin{center}
\subfigure[Time evolution of the radius of the super-critical nucleus.]
{
\includegraphics[width=0.75\linewidth]{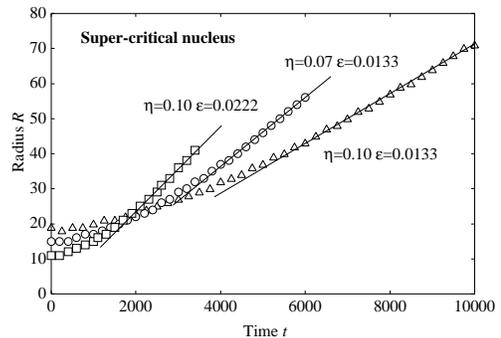}
\label{fig:3a}}
\subfigure[Time evolution of the sub-critical nucleus.]
{
\includegraphics[width=0.75\linewidth]{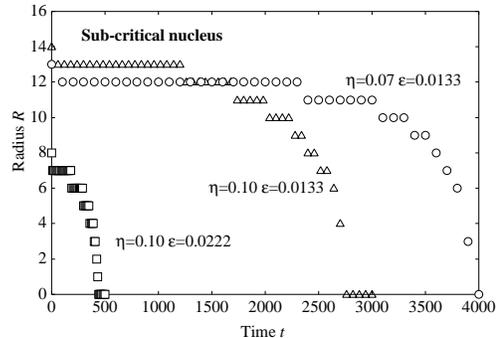}
\label{fig:3b}}
\end{center}
\caption{
(a) The time evolution of the radius of the growing super-critical nucleus for the three sets of potential parameters in Table~\ref{tab:1}.    (b) Time evolution of the radius of the shrinking sub-critical nucleus. }
\label{fig:3}
\end{figure}

The time evolution of the radius $R$ of the nucleus defined as the distance from the center of the nucleus to the position of the cell with $\psi(R)\simeq 0.5$ shows an almost straight line as a function of time expected from the solution of the differential equation Eq.~(\ref{eq:1-13}) for the super-critical growing nucleus ($R>R_{c}$) as shown in Fig.~\ref{fig:3}(a). The velocity $v$ deduced from the cell-dynamics simulation cannot be compared with the theoretical prediction from Eq.~(\ref{eq:1-14}) as the time scale $\tau$ in Eq.~(\ref{eq:1-1}) is unknown.  The same problem of the time scale has already been noticed by the author~\cite{Iwamatsu2005a}.  However, we observe from Table~\ref{tab:1} that the time scale $\tau$ is almost constant and is around $\tau=2.5-2.8$. 

In contrast to the super-critical nucleus, the radius of the shrinking sub-critical nucleus shown in Fig.~\ref{fig:3}(b) does not decrease linearly in time.  The radius remains constant for some periods, then it starts to decrease gradually.  As the radius is getting smaller and smaller it decreases more rapidly.  This non-linear behavior is due to the capillary pressure: The nucleus will be pushed inward by the curvature correction to the pressure. Since we use the discrete cell dynamics, we also observe the staircase structure in Fig.~\ref{fig:3}(b). 

\begin{figure}[htbp]
\begin{center}
\subfigure[Time evolution of the free energy for the super-critical nucleus.]
{
\includegraphics[width=0.75\linewidth]{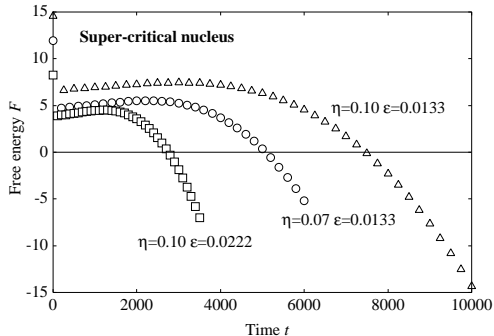}
\label{fig:4a}}
\subfigure[Time evolution of the free energy for the sub-critical nucleus.]
{
\includegraphics[width=0.75\linewidth]{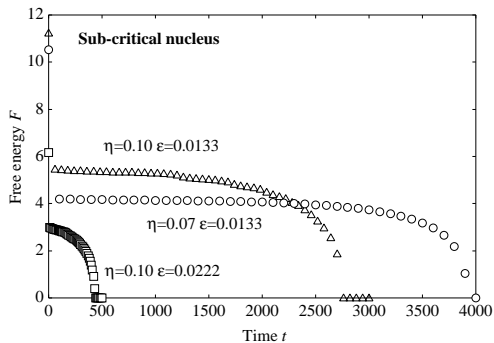}
\label{fig:4b}}
\end{center}
\caption{
(a) The free energy $F$ of the growing super-critical nucleus as the function of time $t$.  (b) The free energy of the shrinking sub-critical nucleus. Both these free energies decrease monotonically in accordance with the theoretical prediction in Eq.~(\ref{eq:1-3}) except at an early stage of growth. 
}
\label{fig:4}
\end{figure}

Similarly, the time evolution of the total free energy given by Eq.~(\ref{eq:1-2}) of the sub- and the super-critical nucleus can be calculated using the temporal order parameter profile $\psi(t,n)=\psi(x,y)$ where $n=(x,y)$ is the site index in two dimension.  The necessary gradients are approximated by the finite differences
\begin{eqnarray}
\frac{\partial \psi}{\partial x}& \rightarrow & \frac{\psi(x+1,y)-\psi(x-1,y)}{2}, \nonumber \\
\frac{\partial \psi}{\partial y}& \rightarrow & \frac{\psi(x,y+1)-\psi(x,y-1)}{2}. 
\label{eq:1-28}
\end{eqnarray}
Figure~\ref{fig:4} shows the time evolution of the total free energy $F$ defined by Eq.~(\ref{eq:1-2}) as the function of the time $t$.  The free energy of the growing super-critical nucleus in Fig.~\ref{fig:4}(a) and that of the shrinking sub-critical nucleus in Fig.~\ref{fig:4}(b) decrease monotonically as the functions of time according to the prediction of Eq.~(\ref{eq:1-3}) except at the early stage of growth in Fig.~\ref{fig:4}(a).  A small increase in the free energy of the super-critical nucleus is probably due to the numerical errors of the coarse-grained derivative in Eq.~(\ref{eq:1-28}).  The initial nucleus at $t=0$ has a much high energy which is indicated by an isolated symbol at $t=0$ in Fig.~\ref{fig:4} because it has an artificial shape with a step-function interface which gives an unrealistically high surface tension from Eq.~(\ref{eq:1-18}).

\begin{figure}[htbp]
\begin{center}
\includegraphics[width=0.8\linewidth]{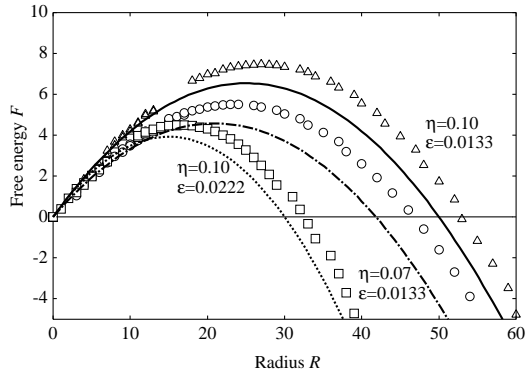}
\caption{
The free-energy barrier along the minimum free-energy path (MFEP) of nucleation deduced from Figs.~\ref{fig:3} and \ref{fig:4}.  The solid lines are the CNT predictions from Eq.~(\ref{eq:1-20}).
}
\label{fig:5}
\end{center}
\end{figure}

Combining Figs.~\ref{fig:3} and \ref{fig:4}, we can deduce the total free energy $F$ of nucleus as the function of its radius $R$ shown in Fig.~\ref{fig:5}.  Since the locus of the phase-field equation Eq.~(\ref{eq:1-1}) in the phase space always follows the route to lower the free energy $F$ of a single nucleus, the curve in Fig.~\ref{fig:5} represents the {\it minimum free-energy path} (MFEP) of nucleation. We have also plotted the MFEP calculated from Eq.~(\ref{eq:1-20}) of the classical nucleation theory (CNT) in Fig.~\ref{fig:5}.  The overall shapes of the MFEP from our simulations and those from the CNT are similar.  In contrast to the dynamical critical radius $R_{c}$ shown in Table~\ref{tab:1}, the position of the thermodynamic critical radius $R_{c}$ estimated from Fig.~\ref{fig:5} of our simulation is almost the same as that from the theoretical predictions Eq.~(\ref{eq:1-23}) based on the CNT.  

However, the absolute magnitude of the free energy from our simulation is always higher than that from the CNT prediction.  It is well recognized that the CNT prediction is incorrect near the spinodal where the free-energy barrier should vanish while the barrier from the CNT erroneously remains finite~\cite{Oxtoby1988,Zeng1991,Iwamatsu1993}.  Since the phase-field model is correct near the spinodal, the free energy from our simulation is expected to be lower than the CNT prediction at least near the spinodal.  Since we are closer to the binodal than the spinodal, the free-energy barriers from DFT could be higher than the CNT predictions. In fact, many other theoretical approaches seem to predict the same trend as ours near the binodal.  For example, Oxtoby and Evans~\cite{Oxtoby1988} predicted that the free energy calculated from their non-classical DFT for the Yukawa fluid can be higher than the CNT prediction near the binodal.  Within the framework of the square-gradient model for the liquid-vapor nucleation~\cite{Iwamatsu1993}, the present author has demonstrated analytically using a double-parabolic free energy that the asymmetry of the free energy between the gas and the liquid phases accounts for the asymmetric behavior of the effective surface tension of the nucleus and, therefore, of the free energy barrier. In this liquid-vapor nucleation model~\cite{Iwamatsu1993}, the free energy barrier for the vapor bubble is always lower than the CNT prediction while the barrier for the liquid droplet can be higher than the CNT prediction near the binodal. We anticipate that a similar asymmetry in our model free energy Eq.~(\ref{eq:1-4}) will account for the free energy barrier higher than the CNT prediction in Fig.~\ref{fig:5}.

Despite the ambiguity in the definition of the radius $R$ of the diffuse interface in the phase-field model, the overall shape of the free-energy barrier in Fig.~\ref{fig:5} is similar to the one from the CNT as well as to the results of other researchers~\cite{Wolde1998,Auer2001,Lutsko2008b}.  Naturally, no anomalous instability or catastrophe which is observed in the $(N,\lambda)$ cluster model~\cite{Weakliem1993} for the bubble nucleation~\cite{Uline2007} as well as for the droplet nucleation~\cite{Uline2008} is observed in our simulation as we are moving along the MFEP of free-energy surface $F$ for a single nucleus.  Since our MFEP is expected to trace the route along the valley on the free-energy surface $F$ guided by the phase-field equation Eq.~(\ref{eq:1-1}), we cannot answer questions regarding the stability of the MFEP against the thermal fluctuation and the effects of the depletion of materials to be incorporated into a growing nucleus.  The latter should be explored using Eqs.~(\ref{eq:1-3a}) or (\ref{eq:1-3b}) instead of Eq.~(\ref{eq:1-1}).  

The $(N,\lambda)$ cluster model is an attempt to count a fluctuation of a nucleus around the MFEP that represents a more condensed or expanded droplet, for example, in the vapor to liquid nucleation.   In the $(N,\lambda)$ two-dimensional space, no instability is observed along the $N$-axis and the instability occurs only along the $\lambda$-axis when the size $\lambda$ is changed while the number of molecule is fixed~\cite{Corti2009}.  The MFEP is the specific route on the $(N,\lambda)$ two-dimensional space on which the size of cluster $\lambda$ is always optimized to the given number of molecule $N$ to make the free energy minimum.

In real nucleation phenomena in a conserved system, such as the condensation or the cavitation in the liquid-vapor nucleation, the evolution of the nucleus does not necessarily follow the MFEP on the free-energy surface $F$ predicted from the phase-field equation Eq.~(\ref{eq:1-1}) for the non-conserved order parameter.  Rather the dynamics will be governed by Eq.~(\ref{eq:1-3a}) for the conserved order parameter or Eq.~(\ref{eq:1-3b}) of the dynamical DFT when the order parameter is density.  The evolution of the nucleus in the conserved system may not follow the MFEP on the free-energy surface $F$ but will follow another route which also guarantees a monotonically decreasing free energy $F$ from the inequality similar to Eq.~(\ref{eq:1-3}).  Then the criticism that is raised by Lutsko~\cite{Lutsko2008a} to the $(N,\lambda)$ cluster model~\cite{Weakliem1993,Talanquer1994,Uline2007} for the conserved liquid-vapor nucleation system is not correct not only mathematically~\cite{Iwamatsu2009} but also physically.  The $(N,\lambda)$ cluster model may include the essential feature of real nucleation phenomena governed by Eqs.~(\ref{eq:1-3a}) or (\ref{eq:1-3b}) for the conserved system.  

In fact, the real picture of nucleation for the conserved system such as the liquid-vapor nucleation will be following: Initially a sub-critical nucleus shrinks or a super-critical nucleus grows according to the kinetics governed by the non-conserved phase-field equation Eq.~(\ref{eq:1-1}) as the depletion of material (monomers) which will be incorporated into the growing nucleus can be neglected. In later stage, due to the depletion of material, the dynamics is now switched from the non-conserved equation in Eq.~(\ref{eq:1-1}) to the conserved one in Eq.~(\ref{eq:1-3a}) or to the dynamical DFT equation in Eq.~(\ref{eq:1-3b}).  For the conserved system, the minimum free-energy path (MFEP) is merely a virtual path on the free-energy surface $F$ that does not take into account the depletion of material due to the increase of the size of nucleus itself or the effect of the other growing nuclei within the same system.  The switching from the non-conserved to the conserved dynamics occurs when this depletion becomes effective.  The distance of this switching point from the saddle point of the MFEP depends on many parameters of the system considered such as the diffusion length, system size, and the number of nuclei presented etc.   The study of the MFEP~\cite{Lutsko2008a,Iwamatsu2009} alone cannot answer this question. This switching point must also be close to the termination point of free-energy surface discovered in the $(N,\lambda)$ cluster model by Uline and Corti~\cite{Uline2007,Uline2008}. Therefore the MFEP could be meaningful in the conserved system only at the early stage of nucleation for the sub-critical nucleus and, probably, only for the super-critical nucleus just over the size of the critical nucleus at the saddle point.  Of course, in the non-conserved system where the order parameter represents, for example, the crystallinity of the solid-liquid phase transition, the MFEP will be totally meaningful up to the time when the nuclei start to coalesce~\cite{Iwamatsu2005a,Iwamatsu2008a}.  Then the KJMA (Kolmogorov-Johnson-Mehl-Avrami) picture of phase transformation~\cite{Christian1965} will be valid in the non-conserved system.  

Finally, it must be noted that the MFEP is usuful only to characterize the nucleation and the evolution of a single nucleus.  The nucleation in real materials may not simply be characterized by the growth of a compact isolated single nucleus.  In fact, Shen and Debenedetti~\cite{Shen1999} have studied the bubble nucleation in the superheated Lennard-Jones fluid, and suggested simultaneous nucleation and coalescence that leads to a ramified non-compact nucleus~\cite{Shen1999}.  Non isothermal effect due to the local temperature fluctuation is also sggested to affect the nucleation process of bubble recently~\cite{Wang2009}. In such a case, we have to consider the heat dissipation equation coupled with the phase field equations, Eqs.~(\ref{eq:1-1}), (\ref{eq:1-3a}) or (\ref{eq:1-3b}) as those equations are considered to describe the isotheraml process.

\section{Conclusion}
\label{sec:sec4}

In conclusion, we have proposed a novel mathematical method to explore the minimum free-energy path (MFEP) on the free-energy surface of the homogeneous nucleation using the phase-field equation for the non-conserved order parameter. By employing the local square-gradient density functional and by solving the phase-field equation for a single circular nucleus using the cell-dynamics method, we can calculate the time evolution of the size, shape, and free energy of the sub- and the super-critical nucleus.  This information is used to extract the MFEP of the homogeneous nucleation.  The method is based on the principle in Eq.~(\ref{eq:1-3}) that the evolution driven by the phase-field equation for the non-conserved order parameter always occurs along the path on the free-energy surface along the direction to lower the free energy.  More rigorous dynamical DFT~\cite{Marconi1999,Archer2004} or the phase-field equation for the conserved order parameter that can also satisfy an inequality similar to Eq.~(\ref{eq:1-3}) may not be effective to prove MFEP though they can be used to study the real picture of nucleation and growth.  Very recently a strategy similar to ours to use the over-damped dynamics in the function space was proposed by Qiu et al.~\cite{Qiu2008} where  an equation similar to the phase-field equation Eq.~(\ref{eq:1-1}) is used as a virtual dynamical equation for the discretized free energy to explore the MFEP of the nucleation for the capillary condensation.

\begin{acknowledgments}
I am indebted to Professor David S. Corti (Purdue University) for drawing my attention to the MFEP problem and for his helpful comments on a draft version of the manuscript.  
\end{acknowledgments}


\begin{thebibliography}{99} 
\bibitem{Oxtoby1992} D. W. Oxtoby, in {\it Fundamentals of inhomogeneous fluids}, ed by D. Henderson, (Marcel Dekker, New York, 1992) Chapeter 10.
\bibitem{Oxtoby1988} D. W. Oxtoby and R. Evans, J. Chem. Phys. {\bf 89}, 7521 (1988).
\bibitem{Zeng1991} X. C. Zeng and D. W. Oxtoby, J. Chem. Phys. {\bf 94}, 4472 (1991). 
\bibitem{Iwamatsu1993} M. Iwamatsu, J. Phys.: Condense. Matter {\bf 5}, 7537 (1993).
\bibitem{Weakliem1993} C. L. Weakliem and H. Reiss, J. Chem.  Phys. {\bf 99}, 5374 (1993).
\bibitem{Talanquer1994} V. Talanquer and D. W. Oxtoby, J. Chem. Phys. {\bf 100}, 5190 (1994).
\bibitem{Uline2007} M. J. Uline and D. S. Corti, Phys. Rev. Lett. {\bf 99}, 076102 (2007).
\bibitem{Reguera2003} D. Reguera, R. K. Bowles, Y. Djikaev, and H. Reiss, J. Chem. Phys. {\bf 118}, 340 (2003).
\bibitem{Gunther2003} L. Gunther, Am. J. Phys. {\bf 71}, 351 (2003).
\bibitem{Wolde1998} P. R. Ten Wolde and D. Frenkel, J. Chem. Phys. {\bf 109}, 9901 (1998).
\bibitem{Auer2001} S. Auer and D. Frenkel, Nature {\bf 409}, 1020 (2001).
\bibitem{Lutsko2008a} J. F. Lutsko, Europhys. Lett. {\bf 83}, 46007 (2008).
\bibitem{Lutsko2008b} J. F. Lutsko, J. Chem. Phys. {\bf 129}, 244501 (2008).
\bibitem{Pusztai2008} T. Pusztai, G. Tegze, G. I. T\'oth, L. {K\"ornyei}, G. Bansel, Z. Fan, and L. Gr\'an\'asy, J.\ Phys.:\ Condens.\ Matter {\bf 20}, 404205 (2008).
\bibitem{Iwamatsu2008a} M. Iwamatsu, J.\ Chem.\ Phys. {\bf 128}, 084504 (2008).
\bibitem{Bagdassarian1994} C. K. Bagdassarian and D. W. Oxtoby, J. Chem. Phys. {\bf 100}, 2139 (1994).
\bibitem{Wild1997} R. Wild and P. Harrowell, Phys. Rev. E {\bf 56}, 3265 (1997).
\bibitem{Chan1977} S-K. Chan, J.\ Chem.\ Phys. {\bf 67}, 5755 (1977).
\bibitem{Langer1992} J. S. Langer, in {\it Solids Far From Equilibrium}, edited by C. Godr\`eche (Cambridge UP, Cambride, 1992), chap. 3.
\bibitem{Qiu2008} C. Qiu, T. Qian, and W. Ren, J. Chem. Phys. {\bf 129}, 154711 (2008).
\bibitem{Marconi1999} U. M. B. Marconi and P. Tarazona, J. Chem. Phys. {\bf 110}, 8032 (1999).
\bibitem{Archer2004} A. J. Archer and R. Evans, J. Chem. Phys. {\bf 121}, 4246 (2004).
\bibitem{Jou1997} H-J. Jou and M. T. Lusk, Phys. Rev. B {\bf 55}, 8114 (1997).
\bibitem{Iwamatsu2005a} M. Iwamatsu and M. Nakamura, Jpn.\ J.\ Appl.\ Phys.\ Part 1 {\bf 44}, 6688 (2005).
\bibitem{Oono1988} Y. Oono and S. Puri, Phys.\ Rev.\ A {\bf 38}, 434 (1988).
\bibitem{Puri1988} S. Puri and Y. Oono, Phys.\ Rev.\ A {\bf 38}, 1542 (1988).
\bibitem{Ren2001a} S. R. Ren and I. W. Hamley, Macromolecules {\bf 34}, 116 (2001)
\bibitem{Christian1965} J. W. Christian, {\it The Theory of Transformations in Metals and 
Alloys} (Pergamon Press, Oxford, 1965). 
\bibitem{Teixeira1997} P. I. C. Teixeira and B. M. Mulder, Phys.\ Rev.\ E {\bf 55}, 3789 (1997).
\bibitem{Uline2008} M. J. Uline and D. S. Corti, J. Chem. Phys. {\bf 129}, 234507 (2008).
\bibitem{Corti2009} D. S. Corti, personal communication (7 April 2009).
\bibitem{Iwamatsu2009} M. Iwamatsu, Europhys. Lett. {\bf 86}, 26001 (2009).
\bibitem{Shen1999} V. K. Shen and P. G. Debenedetti, J. Chem. Phys. {\bf 111}, 3581 (1999).
\bibitem{Wang2009} Z-J. Wang, C. Valeriani, and D. Frenkel, J. Phys. Chem. B {\bf 113}, 3776 (2009).

\end{thebibliography}

\end{document}